\documentclass[aps,showpacs,twocolumn]{revtex4}
\usepackage{graphicx}% Include figure files
\usepackage{epstopdf}
\usepackage{dcolumn}% Align table columns on decimal point

\begin{document}

\title{Nuclear masses set bounds on quantum chaos}

\author{Jos\'e Barea\footnote{E-mail: barea@nuclecu.unam.mx}, 
Alejandro Frank\footnote{E-mail: frank@nuclecu.unam.mx},
Jorge G. Hirsch\footnote{E-mail: hirsch@nuclecu.unam.mx}.}
\affiliation{Instituto de Ciencias Nucleares,
Universidad Nacional Aut\'onoma de M\'exico,
Apartado Postal 70-543, 04510 M\'exico, D.F., M\'exico}
\author{Piet Van Isacker\footnote{E-mail: isacker@ganil.fr }}
\affiliation{Grand Acc\'el\'erateur National d'Ions Louds,
BP 55027, F-14076 Caen Cedex 5, France}
\begin{abstract}
It has been suggested that chaotic motion
inside the nucleus may significantly limit
the accuracy with which nuclear masses can be calculated.
Using a power spectrum analysis we show
that the inclusion of additional physical contributions
in mass calculations,
through many-body interactions or local information,
removes the chaotic signal in the discrepancies
between calculated and measured masses.
Furthermore, a systematic application of global mass formulas
and of a set of relationships among
neighboring nuclei to more than 2000 nuclear masses
allows to set an unambiguous upper bound
for the average errors in calculated masses
which turn out to be almost an order of magnitude smaller
than estimated chaotic components.
\end{abstract}

\maketitle

The importance of an accurate knowledge of nuclear masses to
understand the processes occurring in astrophysical phenomena has
been abundantly stressed~\cite{Rol88}. Though great progress has
been made in the challenging task of measuring the mass of exotic
nuclei, theoretical models are still necessary to {\em predict}
their mass in regions far from stability~\cite{Lunn03}. Advances
in the calculation of atomic masses have been hampered by the
absence of an exact theory of the nuclear interaction and by the
difficulties inherent to quantum many-body calculations, so
diverse models which attempt to bring forth the fundamental
physics of the atomic nucleus have been devised. The simplest
approach is that of the liquid-drop model (LDM)~\cite{Wei35}.
It incorporates
the essential macroscopic terms, which means that the nucleus is
pictured as a very dense, charged liquid drop. Including the
discrete character of the nucleons and their basic interactions
requires more sophisticated treatments. The finite-range droplet
model (FRDM)~\cite{Moll95}, which combines the macroscopic effects
with microscopic shell and pairing corrections, has become the
{\em de facto} standard for mass formulas. A microscopically
inspired model has been introduced by Duflo and Zuker
(DZ)~\cite{Duf94,Zuk94,Duf95} with positive results. Finally,
among the mean-field methods it is also worth mentioning the
Skyrme-Hartree-Fock approach~\cite{Gor01,Sto03}. These mass formulas can
calculate and predict the masses (and often other properties) of
as many as 8979 nuclides~\cite{Lunn03}, but it is in general
difficult to match theory and experiment (for all known nuclei) to
an average precision better than about 0.5 MeV~\cite{Lunn03}.
This minute quantity, corresponding to less
than a part in $10^5$ of the mass of a typical nucleus, still
represents a significant fraction of the energy released in
nuclear decays, strongly affecting the extrapolations of
proton and neutron separation energies
required in astrophysical processes~\cite{Rol88,Lunn03}.

Can these deficiencies of the mass formulas have a chaotic origin?
It was recently suggested that there might be an inherent limit to
the accuracy with which nuclear masses can be calculated~\cite{Abe02},
due to the presence of chaotic motion inside the
atomic nucleus~\cite{Boh02}. The nature of quantum chaos is still
an open question and diverse points of view coexist in the
literature~\cite{Gut90,Sto99}.
Classical chaotic behavior has been understood in terms of
deterministic equations whose time evolution, however, has a
sensitive dependence on their initial conditions. A different kind
of unpredictability arises from the quantum nature of atoms, a
behavior codified in Heisenberg's uncertainty principle. Yet, when
these two sources of unpredictable features combine, the result is
very difficult to fathom. It would appear that quantum uncertainty
can smooth the characteristic signals of chaos, but experiments
with Rydberg atoms in strong magnetic fields~\cite{Blu97} and
quantum dots~\cite{Nak04} clearly display its footprints.

The presence of chaotic motion in highly excited nuclear systems
can be gauged through the statistics of the high-lying energy
levels or resonances~\cite{Wig65,Boh86,Meh90} and their comparison
with Random Matrix Theory (RMT)~\cite{Boh85}. 
It has been convincingly shown that the power spectrum
(Fourier transform squared)
of the fluctuations of a chaotic quantum energy spectra
is characterized by $1/f$ noise~\cite{Rel02}.
The basic corollary of RMT~\cite{Meh90}
is that near or around the neutron absorption
threshold, predictability is hopeless and only a statistical
analysis makes sense. The possible  existence of a remnant of
chaos at the level of the ground state~\cite{Abe02} was addressed
by Bohigas and Leboeuf~\cite{Boh02} from a novel perspective. The
errors among experimental and calculated masses in~\cite{Moll95}
are interpreted in terms of two types of contributions. The first
one is associated with a regular part, related to the underlying
collective dynamics (LDM), plus a shell-energy correction, while
the other is assumed to arise from some inherent dynamics,
possibly higher-order interactions among nucleons~\cite{Boh02},
that lead to chaotic behavior. 

In this context it is worth to analyze the differences
between measured and calculated masses using well-known mass models, looking for signals of quantum chaos and, if these are found,
to figure out if such chaos is inherent to nuclear masses
or an artifact of the approximations in the many-body theories.
It is our purpose to also test the assertion that a lower limit
on how accurately nuclear masses can be calculated
may have already been reached~\cite{Abe02}.
To estimate the best available accuracy,
we use a different approach to calculate known nuclear masses. 
Besides the ``global'' formulas
of which FRDM has become the standard, there are a number of
``local'' mass formulas. These local methods are usually effective
when we require the calculation of the masses of a set of nuclei,
which are fairly close to other nuclei of known mass, exploiting
the relative smoothness of the masses $M(N,Z)$ as a function of
neutron ($N$) and proton ($Z$) numbers to deduce systematic
trends. Among these methods there are the equations connecting the
masses of particular neighboring nuclei known as the Garvey-Kelson (GK)
relations~\cite{Gar66,Gar69}.
These relations do not involve free parameters
and can be {\em derived}
from a simple nuclear-model picture.
Strictly speaking,
they do {\em not} yield an independent calculational tool,
but they do provide strong indication
that a large fraction of the mass values
have a smooth and regular behavior.
In this interpretation, the GK relations can be viewed
as a simple methodology
to estimate nuclear masses from those of its neighbors.

In this Letter we carry out a systematic analysis
of mass errors using different global mass formulas,
supplemented with a study of the estimates
arising from the GK relations
which we use to set an upper limit
on the intrinsic precision
with which masses can be estimated
when enough local information is available.

There are different types of GK relations~\cite{Com88},
and the two best known are~\cite{Gar66,Gar69,Com88}
\begin{eqnarray}
\lefteqn{
-M(N\!+\!1,Z\!-\!2)\!+\!M(N\!+\!1,Z)\!-\!M(N\!+\!2,Z\!-\!1)}
\label{GK1}\\
&&+M(N\!+\!2,Z\!-\!2)\!-\!M(N,Z)\!+\!M(N,Z\!-\!1)=0,
\nonumber\\
\lefteqn{
M(N\!+\!2,Z)\!-\!M(N,Z\!-\!2)\!+\!M(N\!+\!1,Z\!-\!2)}
\label{GK2}\\
&&-M(N\!+\!2,Z\!-\!1)\!+\!M(N,Z\!-\!1)\!-\!M(N\!+\!1,Z)=0.
\nonumber
\end{eqnarray}
These simple equations are based
on the independent-particle shell model
and, furthermore, constructed such
that neutron-neutron, neutron-proton,
and proton-proton interactions cancel.
Both GK relations provide
an estimate for the mass of a given nucleus
in terms of five of its neighbors.
This calculation can be done in six different forms,
as we can choose any of the six terms in the formula
to be evaluated from the others.
Using both formulas, we can have a maximum of 12 estimates
for the mass of a given nucleus,
if the masses of all the required neighboring nuclei are known.
Of course, there are cases
where only 11 evaluations are possible, and so on.
%These sets of nuclei are denoted by GK12, GK11, down to GK1.
About half of all nuclei with measured masses~\cite{Aud03}
can be estimated in 12 different ways
and, in all cases, our estimate corresponds to the average value.
To our knowledge, the systematic application of GK relations in this extended fashion is new. 

Since from this simple method we obtain a specific mass value
entirely determined by the behavior of the neighboring nuclei,
we may conclude that these values 
represent a uniform and smooth component of the nuclear masses.
In this sense they provide an upper estimate
of the intrinsic precision limit
with which they can be calculated,
when enough local information is available.
Fluctuations on top of the regular behavior
predicted by different mass formulas
and the GK relations are analyzed
in the second part of this paper.

In what follows we compare the mass deviations found
in three of the global methods (LDM, FRDM, DZ)
and those in the GK estimates made above.
The corresponding root-mean-square (rms) deviations for $N$ nuclei
\begin{equation}
\sigma_{\rm rms}=
\left[{\frac 1 N}
\sum_{j=1}^N
\left(M_j^{\rm exp}-M_j^{\rm th}\right)^2
\right]^{1/2},
\end{equation}
are displayed in Table~\ref{rms},
which also shows deviations
for the smaller samples GK-$n$
which involve the application of at least $n$ GK relations.
\begin{table}
\caption{The mass rms deviations $\sigma_{\rm rms}$,
in keV, for LDM, FRDM, DZ, and different GK calculations.}
\label{rms}
\begin{tabular}{cccccccc}
\hline
&LDM   &FRDM   &DZ
&GK-1 &GK-4 &GK-7 &GK-12\\
\hline
$A\ge16$&3211 &653 &362 &171 &131 &112 &86\\
$A\ge60$&3177 &529 &320 &115 &100 &88  &80\\
\hline
\end{tabular}
\end{table}
Note the decrease in the errors from LDM to DZ,
as a consequence of the inclusion of shell corrections (FRDM)
and terms that mimic two-body interactions (DZ).
A similar effect is also seen in the GK estimates
with an increase in precision,
proportional to the number of GK relations applied.
We conclude from this analysis that,
if an intrinsic precision limit exists at all
in the calculation of nuclear masses,
it is smaller than 100~keV.
We remark that the uncertainty range implied by these calculations
is consistent with recent estimates of the effect
of statistical fluctuations of high-lying configurations
near the neutron threshold
on the ground-state energy~\cite{Mol04}.
We also stress that experimental errors are lower
than the estimates shown in Table I.
For example, the average experimental error
is $\sigma_{\rm expt}\approx20$~keV
for the nuclei in GK-12. 

Another relevant question is
whether nuclear masses far from stability
can be predicted with a similar accuracy.
In this respect, the GK relations
cannot by themselves offer a simple answer,
but large-scale shell model calculations
shed some light on this important issue~\cite{Hon02}.
Sophisticated new calculations indicate
that predictive power seems robust
against long-distance extrapolations.
Masses of 67 light nuclei in the $fp$ shell, many of them unstable,
were calculated with an average error of 215~keV~\cite{Cau99}.
Errors come mostly from isospin violation,
and do not increase far from stability.
Large-scale shell model calculations
for the $N=126$ Po-Pu isotones~\cite{Cau03}
find errors in the binding energies smaller than 50~keV
along the whole chain, with no increase for unstable nuclei,
and ``imply a high predictive power
for ground-state binding energies
beyond the experimentally known nuclei.''

We now proceed to consider the presence of quantum chaos
and analyze to this end the statistical characteristics
of the mass fluctuations present in the different models. 
A direct 2D study in the $N$-$Z$ plane 
exhibits clear evidence of the correlations
between mass errors in neighboring nuclei~\cite{Hir04}
which, proceeding through the four models (LDM, FRDM, DZ, and GK),
decrease in size.
It is, however, difficult to perform a systematic analysis of them
because of the irregular form of the nuclear-data chart,
and because cuts along fixed $N$, $Z$, or $A$ lines
have a small number of nuclei, making it
difficult to extract unequivocal conclusions~\cite{Hir04,Vel03}.
To overcome these difficulties, we organize all nuclei with
measured masses by ordering them in a {\em boustrophedon} single, 1D
list~\cite{Hir04b}, numbered as follows. Even-$A$ nuclei are
ordered by increasing $N-Z$, while odd-$A$ ones follow a
decreasing value of $N-Z$. Figure~$1$ displays the mass
differences plotted against the order number,
for the four sets of calculations.
\begin{figure}%[h!]
  \begin{center}
\includegraphics[width=8.5cm]{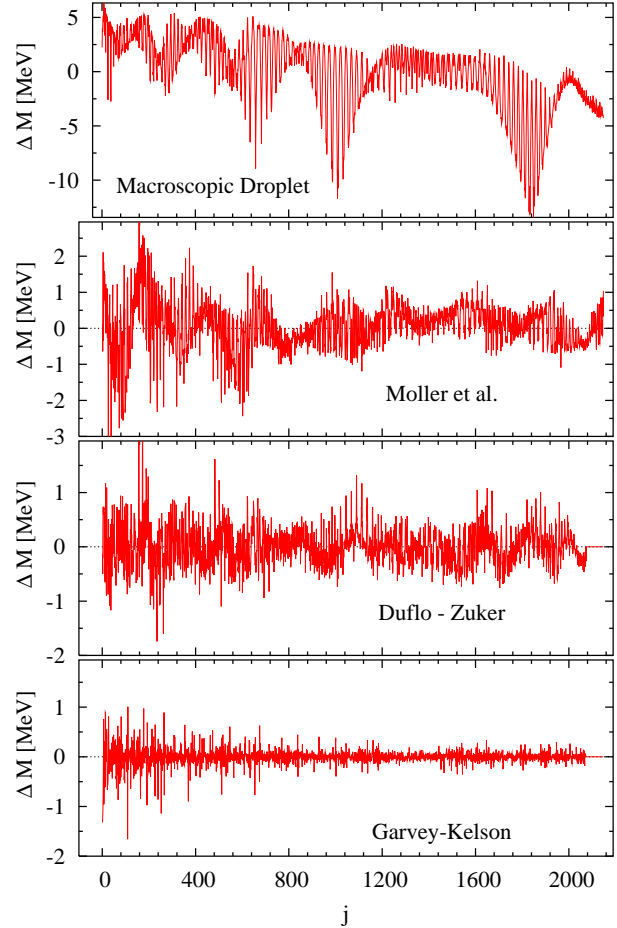}
 \end{center}
\vskip -.7cm
\caption{Differences, in MeV,
between the measured masses~\protect\cite{Aud03}
and those obtained in LDM, FRDM, DZ, and GK,
plotted against the order number $j$.}
\label{difmas-1}
\end{figure}
The LDM errors are conspicuously large around shell closures, as
expected. The presence of strong residual correlations in these
deviations is also apparent from Fig.~$1$. Regions with large
positive or negative errors can be clearly seen in FRDM.
The distribution of the DZ errors is closer to the horizontal axis,
and the correlations are less pronounced,
although not completely absent.
Finally, the remaining GK discrepancies
are small and highly fluctuating,
without any remarkable feature
except for larger errors for $A<60$.

Since the ordering provides single-valued functions,
the discrete Fourier transforms
\begin{equation}
F_k={\frac{1}{\sqrt{N}}}
\sum_j
{\frac{M_j^{\rm exp}-M_j^{\rm th}}{\sigma_{\rm rms}}}
\exp\left({\frac{-2\pi ijk}{N}}\right).
\end{equation}
can be evaluated. The corresponding spectral distributions can
then be fitted to a power law of the form
$|F(\omega)|^2\sim\omega^m$ with $\omega=k/N$.
The squared amplitudes $|F_k|^2$ are shown in Fig.~$2$
where the straight lines are fitted slopes,
which in a log-log plot correspond to the value of $m$
in the corresponding power law.
\begin{figure}%[h!]
  \begin{center}
\includegraphics[width=8.5cm]{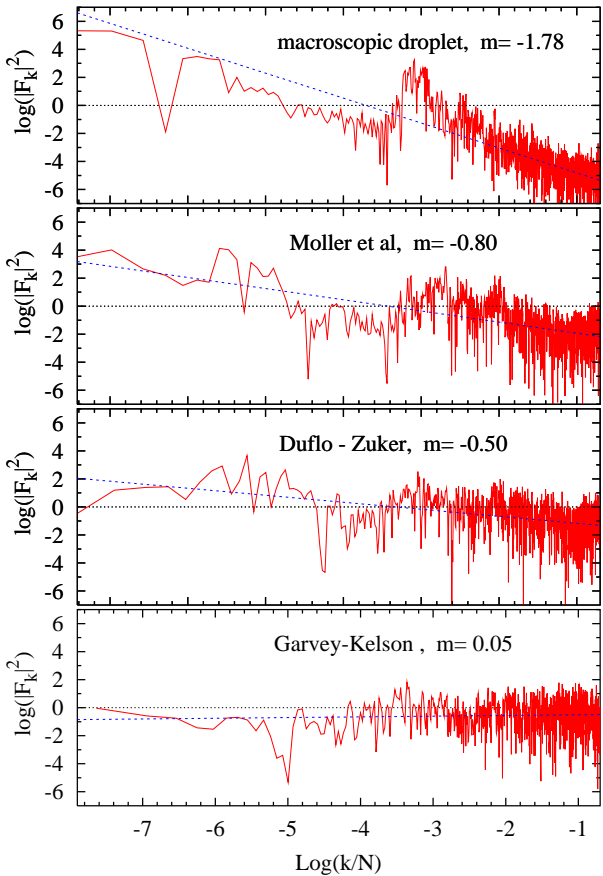}
   \end{center}
\caption{Squared amplitudes
of the Fourier transforms of the mass differences
obtained in LDM, FRDM, DZ, and GK,
plotted against the frequency $\omega=k/N$.}
\label{fourier-l}
\end{figure}
While fluctuations are large, the results are remarkable.
The gradual vanishing of the slope in the different calculations
from top to bottom is evident from the figure.

The fluctuation study allows the application of results well known
from noise analysis. Signals with strong correlations over a long
range tend to have a Brownian motion behavior, where their power
spectrum approach the power law $|F(\omega)|^2 \sim\omega^m$, with
$m=-2$, where $\omega$ is the frequency of the Fourier component.
At the other end of the correlation spectrum is white noise, where
there is no remaining coherence, which implies a flat frequency
dependence $m=0$, that is, all frequencies have equal weight.
Intermediate values of the slope $m$ correspond to transitional
regimes. The first three sets of mass deviations in Fig.~$2$ give
rise to power laws with $m$ ranging from $-1.8$ to $-0.5$,
consistent with a slow transition from substantial correlations to
smaller (but still significant) ones for the associated mass
fluctuations. 
The chaoticity discussed in Ref.~\cite{Boh02},
according to the m =-1 criteria  put forward in Ref.~\cite{Rel02},
seems indeed to be present in the FRDM deviations,
while it tends to diminish in the microscopic DZ calculations.
In contrast, we find that the GK analysis is
essentially consistent with white noise. 

In summary, a careful use of several global mass formulas
and a systematic application of the Garvey-Kelson relations
imply that there is no evidence
that nuclear masses cannot be calculated
with an average accuracy of better than 100~keV.
While mass errors in mean-field calculations like the FRDM
behave like quantum chaos,
with a slope in their power spectrum close to $-1$,
microscopic models' results correspond to smaller slopes.
Finally, for the local GK relations
the remaining mass deviations
behave very much like white noise.
These results seem to confirm
%the conjecture put forward in Ref.~\cite{Boh02}
that the chaotic behavior in the fluctuations
arises from neglected many-body effects.
The more physical information is included in the model,
through many-body terms
or by means of local information
which measure the regular and smooth components
in the mass systematics,
the less presence of chaos is observed.
The local calculations could in principle be reproducing
the chaotic dynamics,
which would imply that no {\it a priori} limit exists
on predictability.
A different interpretation is that the errors
(that simulate chaotic behavior)
are simply removed by the many-body effects.
In any case, our results strongly suggest
an upper bound of approximately 100~keV
as precision limit for the calculation of nuclear masses.
Although the presence of (chaos-related) unpredictability
at such small scale cannot be ruled out by our study,
we believe this is encouraging news 
in the quest for reliable predictions of nuclear masses.

Conversations with R.~Bijker, O.~Bohigas, J.~Dukelsky, J.~Flores, J.M.~Gomez, J. G\'omez-Camacho, P.~Leboeuf, F.~Leyvraz, R.~Molina, W. Nazarewicz, S.~Pittel, A.~Raga, V.~Velazquez, N. Zeldes and A.~Zuker are gratefully acknowledged. This work was supported in part by
Conacyt, Mexico.

\end{document}